\documentclass[letter,twocolumn]{jpsj3}
\usepackage[dvipdfmx]{graphicx}

\makeatletter % bblファイルのエラー回避(Overleaf)
\newcommand\newblock{\hskip .11em\@plus.33em\@minus.07em}
\makeatother

\usepackage{amsmath, amssymb}
\usepackage{bm}
\usepackage{mathrsfs}
\usepackage{physics}
\usepackage{comment}
\usepackage[normalem]{ulem}

\usepackage{color}

\title{Nonlinear Piezomagnetic Effects in $g$-wave Altermagnets}
\author{Yuuki Ogawa and Satoru Hayami}
\inst{Graduate School of Science, Hokkaido University, Sapporo 060-0810, Japan}

\abst{
We theoretically study the generation of net magnetization induced by static strain in $g$-wave altermagnets, which exhibit the symmetric spin-split band structure under collinear spin textures free from the relativistic spin--orbit coupling.
By analyzing a tight-binding model in a two-dimensional tetragonal system, we show that the $g$-wave altermagnets give rise to the nonlinear piezomagnetic effect, where a net magnetization is induced by the second-order strain. 
We compare the results for the $g$-wave altermagnets with those for the $d$-wave altermagnets, where the linear piezomagnetic effect occurs.
As a result, we find that the induced magnetization is enhanced when the Fermi level lies on the band with the large spin splitting in both cases. 
We also show that the magnitudes of the induced magnetization are comparable to each other. 
Our results indicate that the nonlinear piezomagnetic effect is a good phenomenon to characterize the physical properties in $g$-wave altermagnets.
}

\date{\today}
\begin{document}
\maketitle

Piezomagnetic effect (PME), which corresponds to the magnetization generation induced by strain, has been studied since the 1950s~\cite{Tavger1956, Dzialoshinskii1958, Moriya1959, Borovik-Romanov1960, Borovik-romanov1994, Lukashev2008, Arima2013, Zemen2017}.
Owing to the development of technology, observations of the PME have been reported in a variety of materials, such as UO$_2$~\cite{Jaime2017} and Mn$_x$Ni$_{1-x}$F$_2$~\cite{Van_Haren2023}, and recently, it has been detected in Weyl semimetal Mn$_3$Sn at room temperature~\cite{Ikhlas2022}.
Such a PME has been observed in antiferromagnetic (AFM) materials without the time-reversal symmetry, since the piezomagnetic tensor $Q_{ijk}$ corresponds to a third-rank axial $c$ tensor as $M_\mu = Q_{\mu\nu\eta} u_{\nu\eta}$ for $\mu,\nu,\eta=x,y,z$ ($M_\mu$ and $u_{\nu\eta}$ stand for the magnetization and strain, respectively).
More precisely, the PME can occur when either the magnetic dipole, magnetic toroidal quadrupole, or magnetic octupole belong to the identity irreducible representation~\cite{Yatsushiro_PhysRevB.104.054412}, where 66 magnetic point groups are categorized in total~\cite{birss1964symmetry, Dresselhaus_Dresselhaus_Jorio, hayami2024unified}.

Recently, it has been revealed that some collinear AFM materials with zero net magnetization can exhibit momentum-dependent spin-split bands even in the absence of the relativistic spin--orbit coupling (SOC)~\cite{Noda2016, Okugawa2018, Ahn2019, Naka2019, Hayami_YK2019, Hayami_PhysRevB.102.144441, Yuan2020, Naka_PhysRevB.103.125114, Yuan2021PRMat, Yuan2021PRB}.
Since such AFM materials are qualitatively distinct from conventional ferromagnets that exhibit the uniform Zeeman-type spin-split bands or antiferromagnets with the spin-degenerate bands owing to the presence of the effective time-reversal symmetry consisting of the translational and time-reversal operations, they are potentially expected for new spintronic devices without relying on heavy elements with the large SOC~\cite{Ahn2019, Naka2019, Sun_PhysRevB.108.L140408, Bai_PhysRevLett.130.216701, Giil_PhysRevB.110.L140506}.
To date, such AFM materials are called altermagnets (AMs)
~\cite{Gonzalez-Hernandez2021, Smejkal2022PRX1, Smejkal2022PRX3, Smejkal2022PRX4}.
The AMs are categorized according to the rotational symmetry of the spin-split band structures; $d$-wave AMs with the spin splitting $k_\mu k_\nu \sigma$ and $g$-wave AMs with $k_\mu k_\nu k_\eta k_\gamma \sigma$, where $k_\mu$ is the $\mu$ component of the wave vector and $\sigma$ is the spin along the collinear AFM moment direction. 
Microscopically, such an effective coupling between wave vectors and spin is brought about by the interplay between the sublattice-dependent hoppings and AFM mean fields~\cite{Hayami_YK2019, Hayami_PhysRevB.102.144441, Roig_PhysRevB.110.144412}.
These spin-momentum couplings result in a variety of cross correlations, such as the spin current generation~\cite{Ahn2019,Naka2019}, 
PME~\cite{Ma2021, Zhu2024, Aoyama2024, McClarty2024_PhysRevLett.132.176702, Wu2024, Yershov2024, Chen2024}, 
nonlinear magnetoelectric effect~\cite{Oike_PhysRevB.110.184407}, and Coulomb drag~\cite{PhysRevLett.134.136301}, which are free from the relativistic SOC. 
For example, when the spin-split band structure is characterized by the functional form $k_xk_y\sigma$ in $d$-wave AMs, we can infer a linear coupling between the $u_{xy}$-type strain and the magnetization in the direction of collinear spin moments.
Meanwhile, such a linear relation between the strain and magnetization is no longer valid for $g$-wave AMs owing to the higher-order spin-split band structure. 
A similar argument also holds for the spin current generation; the $g$-wave AMs do not show the spin current generation within the linear response theory. 
In this way, physical phenomena characteristic of $g$-wave AMs remain less explored compared to those of $d$-wave AMs, despite the experimental observation of $g$-wave AMs in various materials, such as MnTe~\cite{Krempasky2024, Lee2024, Osumi2024, observedPME} and CrSb~\cite{Ding2024, Reimers2024, Zeng2024}. 
Thus, it is highly desired to clarify physical phenomena driven by $g$-wave AMs.

\begin{figure}
\centering
\includegraphics[width=\linewidth]{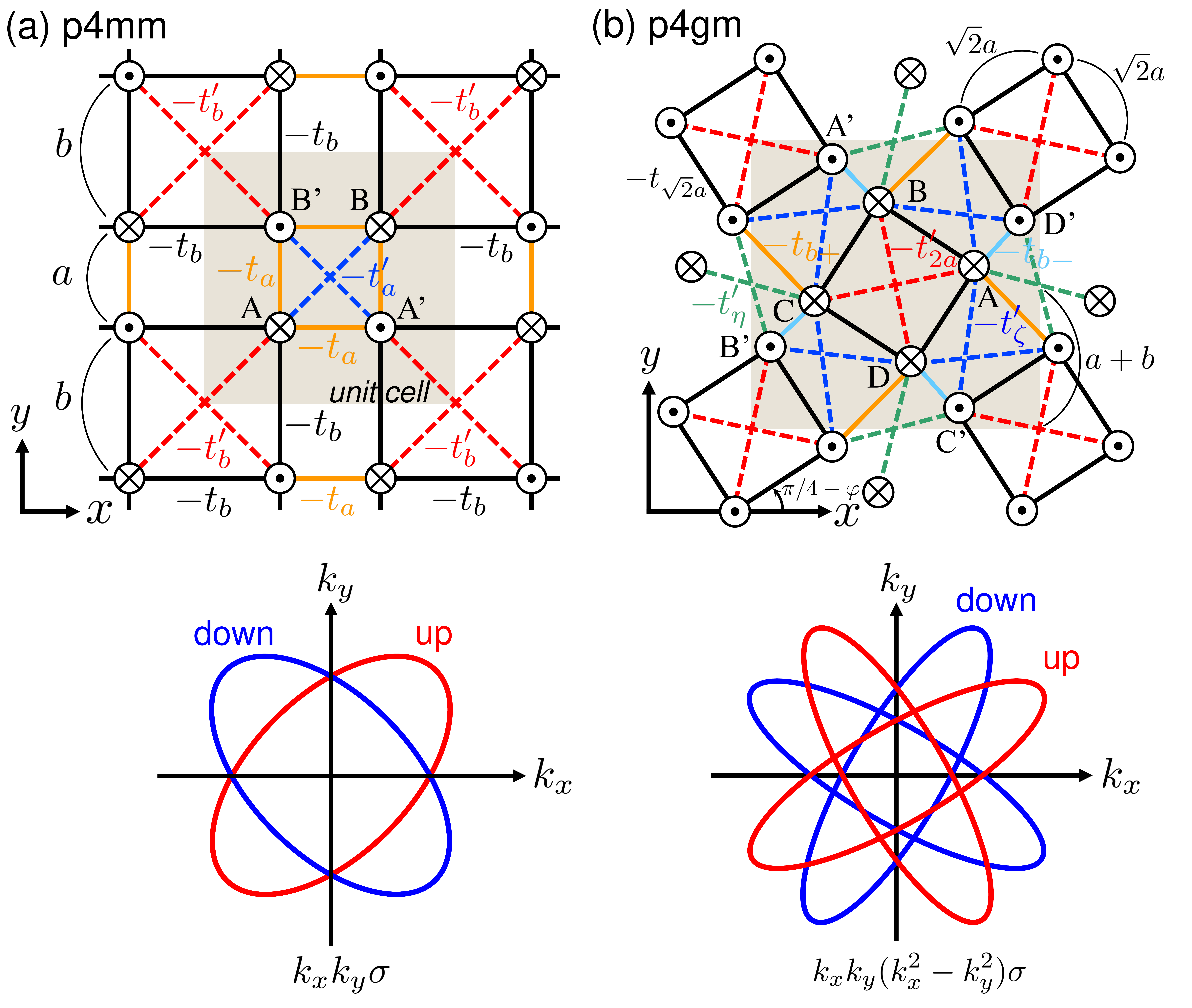}
\caption{
(Color online)
(Top panel) Lattice and magnetic structures under the two-dimensional (a) $p4mm$ and (b) $p4gm$ space group; the former consists of four sublattices (A, B, A', B'), while the latter consists of eight sublattices (A, B, C, D, A', B', C', D'), where the symbols with/without prime represent sublattices with up/down spin moments.
The solid and dashed lines represent the hoppings 
$(-t_a, -t_b)$ [$(-t_{\sqrt{2}a}, -t_{b+}, -t_{b-})$] and $(-t'_a, -t'_b)$ [$(-t'_{2a}, -t'_{\eta}, -t'_{\zeta})$]
in (a) [(b)], respectively. 
The filled square represents the unit cell.
The AFM orderings in (a) and (b) correspond to the $d$-wave and $g$-wave AMs, which results in the symmetric spin-split band structure in the form of $k_x k_y \sigma$ and $k_x k_y (k^2_x-k^2_y)\sigma$ in momentum space, respectively, as shown in the bottom panel.
} 
\label{model}
\end{figure}

In this Letter, we theoretically investigate physical phenomena in $g$-wave AMs by focusing on the nonlinear PME beyond phenomenological approaches~\cite{McClarty2024_PhysRevLett.132.176702}.
We show that the $g$-wave AMs exhibit the nonlinear PME when static uniaxial strain is applied to the low-symmetry directions in the system. 
To demonstrate that, we examine a tight-binding model under a two-dimensional tetragonal lattice structure with the space group $p4gm$. 
By dealing with the effect of lattice distortion as the modulation of hopping integrals, we calculate the net magnetization under the uniaxial tensile and shear strain.
We show that the induced magnetization is proportional to the second-order strain effect, whose magnitude is largely enhanced when the Fermi level is located at the bands with the large spin splitting. 
In addition, we compare the results in $g$-wave AMs with those in the $d$-wave AMs by additionally considering a tight-binding model under the $p4mm$ space group. 
We find that the magnitudes of the induced magnetization are almost comparable to each other, although their amplitude dependence is different from each other. 
We also discuss the dependence on the applied angle of the strain, which reflects the angle dependence of the spin-split band structure.

In order to clarify the relation between $g$-wave ($d$-wave) AMs and nonlinear (linear) PME at the microscopic level, we consider a specific tight-binding model in two-dimensional tetragonal systems with four and eight sublattices, as shown in the top panel of Figs.~\ref{model}(a) and (b), respectively~\cite{Hayami_YK2019}, although the following schemes can be applied to other tight-binding models with different lattice structures when the system satisfies the symmetry of $d$- or $g$-wave AMs.
The former is classified as the $p4mm$ space group and the latter as the $p4gm$ space group, where the $d$-wave and $g$-wave spin splittings in momentum space occur by considering the collinear AFM structures in Figs.~\ref{model}(a) and (b), respectively, as detailed below.
Accordingly, the $p4mm$ system exhibits the spin splitting of $k_xk_y\sigma$, while the $p4gm$ system exhibits that of $k_xk_y(k_x^2-k_y^2)\sigma$, as shown in the bottom panel of Figs.~\ref{model}(a) and (b), respectively. 
The model Hamiltonian is given by
$H = H_{\rm hop} + H_{\rm MF}$,
with
\begin{align}
 H_{\rm hop} 
 &= \displaystyle \sum_{\bm{r}_i,\bm{r}_j} \sum_{\alpha,\beta} 
    E(\bm{r}_{j\beta}-\bm{r}_{i\alpha}) c_{i\alpha}^\dagger c_{j\beta}  ,  
\\
 H_{\rm MF} 
 &= h \displaystyle \sum_{\bm{r}_i} 
    \biggl[ \, \sum_{\gamma} c_{i\gamma}^\dagger \, \sigma^z \, c_{i\gamma} - \sum_{\gamma'} c_{i\gamma'}^\dagger \, \sigma^z \, c_{i\gamma'} \biggr] , 
\end{align}
where a two-component operator, 
$c_{i\alpha} = {}^t (c_{i\alpha\uparrow}, c_{i\alpha\downarrow})$,  
that annihilates an %$s$ 
electron at sublattice 
$\alpha$ (=A, $\cdots,$ B', or A, $\cdots$, D') in the $i$-th unit cell is introduced.
$H_{\rm hop}$ represents the hopping Hamiltonian, where $E(\bm{r}_{j\beta}-\bm{r}_{i\alpha})$ is the hopping integral between the lattice sites at $\bm{r}_{i\alpha}$ and $\bm{r}_{j\beta}$, depending only on the distance between them; 
$\bm{r}_{i\alpha}=\bm{r}_i + \bm{\xi}_\alpha$ 
is the coordinate of the sublattice $\alpha$ 
[$\bm{\xi}_\alpha$ is taken from the center of the unit cell shown by the shaded square region in Figs.~\ref{model}(a) and \ref{model}(b)] in the $i$-th unit cell at $\bm{r}_i$; 
see the Supplemental Material~\cite{suppl} for the details of the model.
The area of the unit cell is given by $(a+b)^2$ for the $p4mm$ model and $4(a+b)^2$ for the $p4gm$ models.
For the $p4mm$ model, there are four kinds of hopping integrals; 
$-t_a$, $-t_b = -(a/b)^2 t_a$, $-t_a'$, and 
$-t_b'= -(a/b)^2t_a'$, 
and we set $t_a$ as a unit of energy scale ($t_a=1$) and $a/b=0.9$.
For the $p4gm$ model, there are six kinds of hopping integrals; 
$-t_{\sqrt{2}a}$, $-t_{b\pm}$, $-t_{2a}'$, $-t_{\eta}'$, and $-t_{\zeta}'$, where
$-t_{b\pm}$ ($-t_{\eta}'$ and $-t'_{\zeta}$) 
is related to $-t_{\sqrt{2}a}$ ($-t_{2a}'$)~\cite{suppl}, 
and we set $t_{\sqrt{2}a}$ as a unit of energy scale ($t_{\sqrt{2}a}=1$), $b/a=0.9$, and $\varphi=0.2$.

$H_{\rm MF}$ represents the AFM mean-field Hamiltonian.
The sum of $\gamma$ (=A, B, or A, B, C, D) and $\gamma'$ (=A', B', or A', B', C', D') represents the sum of sublattices whose collinear spin moments are down and up, as shown in Figs.~\ref{model}(a) and \ref{model}(b).
$\sigma^z$ is the $z$ component of the Pauli matrix in spin space. 
Although the spin direction is arbitrarily taken when the SOC is neglected, we suppose its direction along the out-of-plane direction for simplicity ($\sigma^z \equiv \sigma$). 
We set $h=1$.

\begin{figure}
\centering
\includegraphics[width=\linewidth]{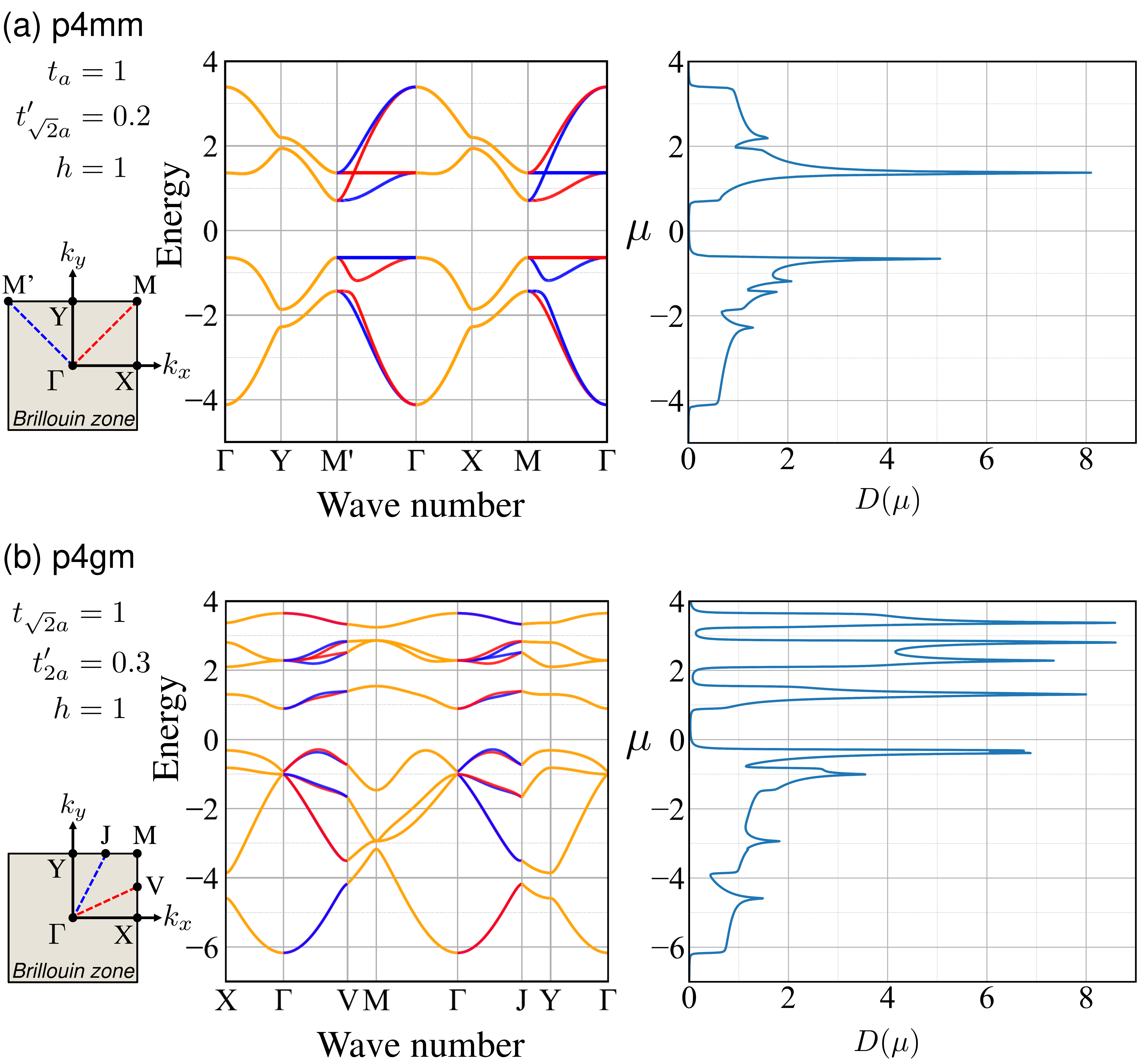}
\caption{
(Color online)
(Left panel) Spin-split band structures along the high-symmetry lines in the Brillouin zone for the (a) $p4mm$ and (b) $p4gm$ models. 
The red, blue, and orange colors stand for the up, down, and zero spin polarizations at each wave number. 
The model parameters and the Brillouin zones are presented on the leftmost side; 
V and J in (b) represent low-symmetric wave numbers, which are defined by 
V $=(\frac{\pi}{2(a+b)}, \frac{\pi}{4(a+b)})$ and J $=(\frac{\pi}{4(a+b)}, \frac{\pi}{2(a+b)})$, respectively. 
(Right panel) The corresponding density of states.   
} 
\label{band_DOS}
\end{figure}

In the presence of the collinear AFM orderings in Figs.~\ref{model}(a) and \ref{model}(b), the symmetric spin-split band structures appear even without the SOC.
We demonstrate the band structures in the $p4mm$ case at $t_a'=0.2$ in Fig.~\ref{band_DOS}(a) and in the $p4gm$ case at $t_{2a}'=0.3$ in Fig.~\ref{band_DOS}(b).
In Fig.~\ref{band_DOS}(a), the symmetric spin splitting occurs in the M-$\Gamma$ and M'-$\Gamma$ lines, while no spin splitting occurs in the other high-symmetry lines. 
This indicates that the functional form of the spin splitting is given by $k_x k_y \sigma$, as schematically shown in the bottom panel of Fig.~\ref{model}(a). 
Meanwhile, in Fig.~\ref{band_DOS}(b), no symmetric spin splitting occurs in all the high-symmetry lines, whereas the spin splitting can be found in the low-symmetry $\Gamma$-V and $\Gamma$-J lines, where V and J correspond to the low-symmetry wave numbers, as shown in Fig.~\ref{band_DOS}(b). 
This indicates the higher-order spin splitting in the form of $k_x k_y (k_x^2 - k_y^2) \sigma$, as schematically shown in the bottom panel of Fig.~\ref{model}(b). 
The appearance of the symmetric spin splittings is consistent with the magnetic symmetry.

Next, we consider the effect of the static strain.
We specifically take into account the lattice distortion by the modulation of hopping integrals given by 
$E(\bm{r}_{j\beta}+\delta\bm{R}(\bm{r}_{j\beta}) - \bm{r}_{i\alpha}-\delta\bm{R}(\bm{r}_{i\alpha}))$, where $\delta\bm{R}(\bm{r})$ is local atomic displacement.
We suppose that the modulation of the hopping integrals is dependent only on the distance between electrons (independent of the direction).
We also assume the inverse-square dependence of the hopping amplitudes on the interatomic distance following the Harrison's rule~\cite{Froyen1979, Harrison1980}.
By performing a similar manner in Ref.~\citen{Ogawa2023}, we evaluate the hopping integrals between the lattice points $\bm{r}_{i\alpha}$ and $\bm{r}_{j\beta}=\bm{r}_{i\alpha}+\bm{R}_{\alpha\beta}$ up to the lowest order of $\delta\bm{R}$, which is given by
\begin{align}
&E(\bm{R}_{\alpha\beta} + \delta\bm{R}(\bm{r}_{j\beta}) -\delta\bm{R}(\bm{r}_{i\alpha}))
\nonumber \\
&=
E(\bm{R}_{\alpha\beta})
\biggl[ 1 - 2\frac{\bm{R}_{\alpha\beta}}{\abs{\bm{R}_{\alpha\beta}}} \cdot \frac{\delta\bm{R}(\bm{r}_{j\beta}) - \delta\bm{R}(\bm{r}_{i\alpha})}{\abs{\bm{R}_{\alpha\beta}}}
\biggr] .
\end{align}

Consequently, the hopping Hamiltonian is expressed as
\begin{align}
\tilde{H}_{\rm hop}
&=
\sum_{\bm{k}}\sum_{\alpha,\beta} 
\biggl[ \, \sum_{\bm{R}_{\alpha\beta}} E(\bm{R}_{\alpha\beta}) e^{i\bm{k}\cdot\bm{R}_{\alpha\beta}} 
\nonumber
\\
& \quad
\Bigl\{ 1 - 2 \left( \ell^2 u_{xx} + 2\ell m u_{xy} + m^2 u_{yy} \right) \Bigr\}
\biggr]
c_{\bm{k}\alpha}^\dagger c_{\bm{k}\beta} \label{eq:Hhop(k)} ,  
\end{align}
where we adopt the momentum-space representation by performing the Fourier transformation from $c_{i\alpha}$ to $c_{\bm k\alpha}$. 
$(\ell , m)$ is the direction cosine of $\bm{R}_{\alpha\beta}$ satisfying $\ell^2 +m^2=1$. 
The list of $\bm{R}_{\alpha\beta}$ and more detailed calculations are given in the Supplemental Material~\cite{suppl}. 
$u_{\nu\eta}=\frac{1}{2}(\pdv{}{x^\nu}\delta R^\eta + \pdv{}{x^\eta}\delta R^\nu)_{\bm{q}} \simeq \frac{1}{2}(iq_\nu \delta R^\eta_{\bm{q}} + iq_\eta \delta R^\nu_{\bm{q}})$, where the displacement is given by $\delta \bm{R}(\bm{r})=\delta \bm{R}_{\bm{q}}e^{i\bm{q}\cdot\bm{r}}$.
For uniaxial tensile (shear) strain, the lattice displacement $\delta \bm{R}$ is parallel (perpendicular) to its wave vector $\bm{q}$.
It is noted that Eq.~(\ref{eq:Hhop(k)}) is valid when a continuum description is adopted in the small limit of the wave vector of the static strain.

For the total Hamiltonian $\tilde{H}_{\rm hop}+H_{\rm MF}$, we evaluate the net magnetization, which is given by
\begin{align}
\langle{M}\rangle 
= 
\frac{1}{N}\sum_{n,\bm{k}} \bra{n\bm{k}}\mathcal{M}\ket{n\bm{k}}f(E_{n\bm{k}}),
\end{align}
where $\mathcal{M}=-\mu_{\rm B}\sum_{\bm{k}}\sum_{\alpha} c_{\bm{k}\alpha}^\dagger \sigma^z c_{\bm{k}\alpha}$ 
with the Bohr magneton $\mu_{{\rm B}}$ and $f(E)=[e^{\beta(E-\mu)}+1]^{-1}$ is the Fermi distribution function with the inverse temperature $\beta$ and the chemical potential $\mu$. 
$N$ is the number of unit cells.
$\ket{n\bm{k}}$ and $E_{n\bm{k}}$ represent the eigenstates and eigenvalues for the $n$-th band, respectively. 
In the following calculations, we set $\mu_{\rm B}=1$ and $\beta=10^3$.

As an input strain, we adopt the uniaxial tensile strain ($\bm{q}\parallel\delta\bm{R}$), which is given by
\begin{align}
u_{xx}=u\cos^2\theta, \ u_{yy}=u\sin^2\theta, \ u_{xy}=\frac{1}{2}u\sin{2\theta},
\end{align}
where $u$ is the strain-field magnitude, approximately equals to the ratio of the displacement magnitude to the lattice constant, indicating the degree of lattice elongation due to the tensile strain, and $\theta=\arctan(q_y/q_x)$ is the angle of the applied tensile strain measured from the $x$-axis.
The symmetry of the system is lowered depending on $\theta$; 
for example in the case of $\theta = \pi/2$ and $u>0$, a uniform elongation along the $y$-axis reduces the space group of the $p4mm$ (or $p4gm$) model to one of its subgroups such as $p2mm$ (or $p2gm$).
The hopping integrals are also modulated according to Eq.~(3) so as to satisfy the subgroup symmetry.
In our calculations, we mainly consider values of $u$ up to $0.1$, which are reasonably close to realistic estimates, although $u$ has typically been estimated to reach approximately 0.002 (0.2 \%) in Mn$_3$Sn~\cite{Ikhlas2022}, for example, and to reach at most 0.05 (5 \%) in V$_2$Se$_2$O~\cite{Ma2021} and Cr$_2$S$_2$~\cite{Chen2024}, based on first-principles calculations.
We also investigate the case of shear strain, which does not qualitatively alter the following results; see the Supplemental Material~\cite{suppl}.

Figures~\ref{M_vs_mu}(a) and \ref{M_vs_mu}(b) show the chemical potential $\mu$ dependence of the magnetization $\langle M \rangle$ when the uniaxial tensile strain with $u=0.1$ are applied for the $p4mm$ and $p4gm$ models, respectively. 
We set $\theta=\pi/4$ in Fig.~\ref{M_vs_mu}(a) and $\theta=\pi/8$ in Fig.~\ref{M_vs_mu}(b).
The results indicate that the magnetization is induced by the uniaxial tensile strain in both models. 
Compared to the spin-split band structure in the left panel of Fig.~\ref{band_DOS}, the induced magnetization is strongly enhanced when the Fermi level lies at the bands with the large spin splitting; for example, in the $p4mm$ model, the magnetization shows the peak structures at $\mu \simeq -0.7$ and $\mu=1.3$ in Fig.~\ref{M_vs_mu}(a), where the large spin splittings along the M'-$\Gamma$ and M-$\Gamma$ lines are found. 
Furthermore, one finds that the density of states $D(\mu)$ is also enhanced at such chemical potentials, which arises from the flat-like spin-split band structure, as shown in Fig.~\ref{band_DOS}(a). 
In other words, there are two important factors in obtaining the large induced magnetization: One is the large spin splitting and the other is the large density of states. 
Such a feature is also found in the $p4gm$ model in Fig.~\ref{M_vs_mu}(b). 
It is noted that the multiple peak structures in $\langle M \rangle$ in Fig.~\ref{M_vs_mu}(b) compared to the peak structures of the density of states in Fig.~\ref{band_DOS}(b) are attributed to the small energy splitting in the presence of the tensile strain as shown in Fig.~\ref{M_vs_mu}(b).

\begin{figure}
\centering
\includegraphics[width=\linewidth]{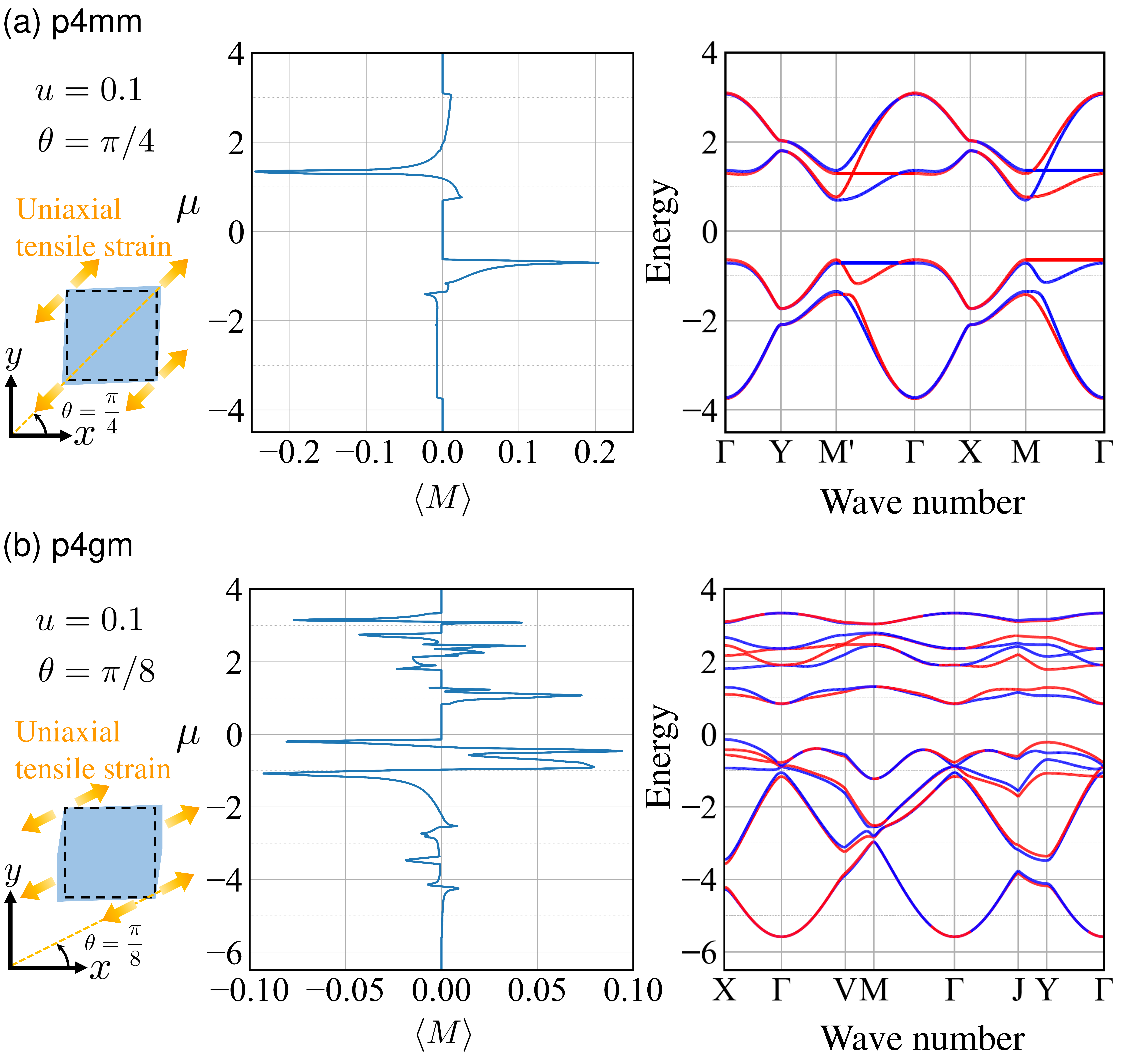}
\caption{
(Color online)
(Left panel) Chemical potential dependence of the net magnetization $\langle M \rangle$ induced by the uniaxial tensile strain for the (a) $p4mm$ and (b) $p4gm$ models. 
(Right panel) The corresponding band structures under the strain, where red (blue) color represents spin-up (down).
} 
\label{M_vs_mu}
\end{figure}

\begin{figure}
\centering
\includegraphics[width=\linewidth]{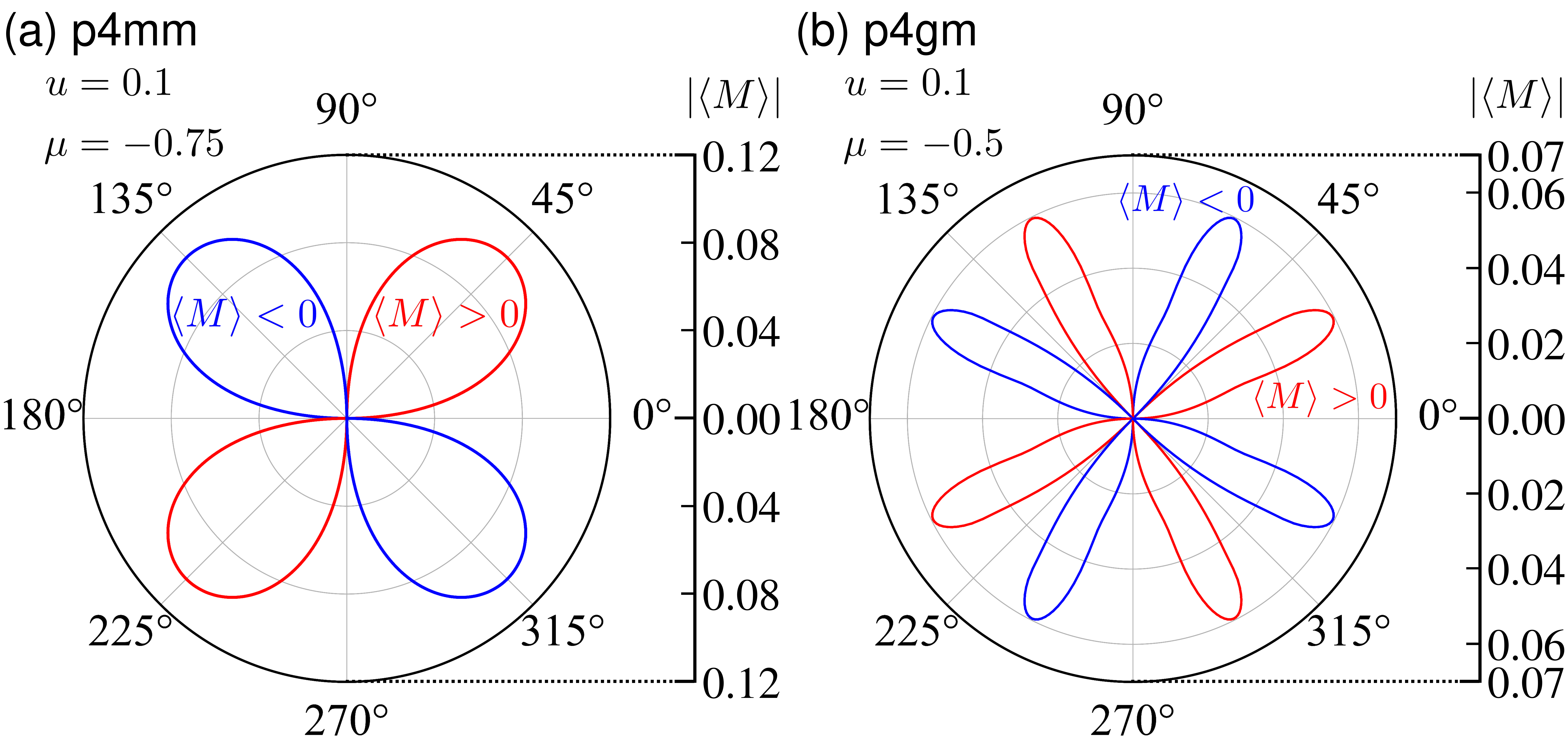}
\caption{
(Color online)
Dependence of the net magnetization induced by the uniaxial tensile strain on the angle of application for the (a) $p4mm$ and (b) $p4gm$ models.
} 
\label{M_vs_theta}
\end{figure}

Next, we show the angle $\theta$ dependence of the induced magnetization in terms of the uniaxial tensile strain at $u=0.1$ in Fig.~\ref{M_vs_theta}.
We set $\mu=-0.75$ in Fig.~\ref{M_vs_theta}(a) and $\mu=-0.5$ in Fig.~\ref{M_vs_theta}(b).
The magnetization behaves as $\sin 2\theta$ for the $p4mm$ model in Fig.~\ref{M_vs_theta}(a) and as $\sin 4\theta $ for the $p4gm$ model in Fig.~\ref{M_vs_theta}(b). 
Such an angle dependence is consistent with the symmetry of the spin-split band structure in the bottom panel of Figs.~\ref{model}(a) and \ref{model}(b).
In other words, no uniform magnetization is induced when the tensile strain is applied in the direction where the symmetric spin splitting is absent; $\langle M \rangle=0$ for $\sin2\theta=0$ in Fig.~\ref{M_vs_theta}(a) and for $\sin 4\theta=0$ in Fig.~\ref{M_vs_theta}(b).

It is noteworthy that the uniform magnetization is induced in the $p4gm$ model, which corresponds to the $g$-wave AM.  
The necessary condition is $\sin 4\theta \neq 0$ in terms of the angle of the input 
tensile strain, which means the input along the low-symmetry direction is important in obtaining the net magnetization. 
This is understood from the symmetry viewpoint, since the low-symmetry tensile strain lowers the crystal symmetry from the tetragonal to monoclinic symmetry so that two electric quadrupoles with $x^2-y^2$ and $xy$ dependence, i.e., $Q_v$ and $Q_{xy}$, belong to the identity irreducible representation, which enables the effective coupling in the form of $Q_v Q_{xy}\sigma$ under the AFM mean field.

\begin{figure}
\centering
\includegraphics[width=\linewidth]{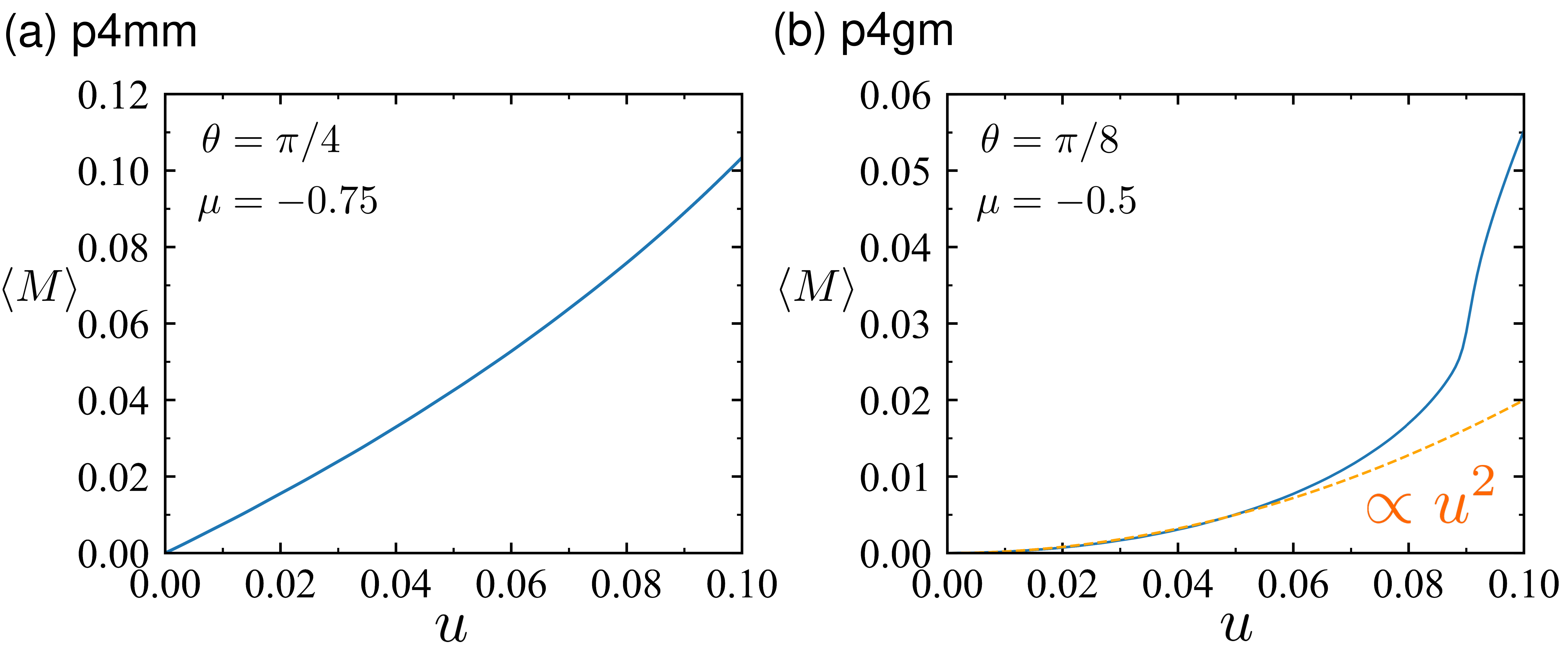}
\caption{
(Color online)
Dependence of the net magnetization induced by the uniaxial tensile strain on the magnitude of the strain field for the (a) $p4mm$ and (b) $p4gm$ models.
The orange dashed line in (b) represents the fitting line.
} 
\label{M_vs_u}
\end{figure}

The above results indicate the difference of the amplitude dependence in terms of the 
tensile strain between the $p4mm$ and $p4gm$ models because of the relation as $\sin 2\theta \propto u_{xy}$ and $\sin4\theta \propto u_{xy} (u_{xx}-u_{yy})$.
Thus, it is expected that the induced magnetization is proportional to $u$ for the $p4mm$ model and $u^2$ for the $p4gm$ model. 
Indeed, such a feature is confirmed as shown in Figs.~\ref{M_vs_u}(a) and \ref{M_vs_u}(b), where the comparable magnitude is obtained for $u \simeq 0.1$. 
In this way, the linear PME occurs in the $d$-wave AM and the nonlinear PME occurs in the $g$-wave AM.

Finally, let us comment on the important model parameters to obtain the linear and nonlinear PME in the present model.
In the case of the $p4mm$ model, the diagonal hoppings, $t_a'$ and $t_b'$, correspond to the essential model parameters. 
This is because the symmetric spin splitting vanishes for $t_a'=t_b'=0$~\cite{Hayami_YK2019}.
Meanwhile, in the case of the $p4gm$ model, the combinations of at least three hoppings, such as 
$(-t_{\sqrt{2}a},-t_{b+},-t'_\eta)$, 
$(-t_{\sqrt{2}a},-t_{b-},-t'_\eta)$, 
$(-t_{b+},-t'_{2a},-t'_\eta)$, 
$(-t_{b+},-t'_{2a},-t'_\zeta)$, 
$(-t_{b+},-t_{b-},-t'_{2a})$, 
$(-t_{b+},-t'_\eta,-t'_\zeta)$, 
$(-t_{b-},-t'_{2a},-t'_\eta)$, 
$(-t_{b-},-t'_{2a},-t'_\zeta)$, 
and $(-t_{b-},-t'_\eta,-t'_\zeta)$ 
are important in inducing the magnetization as well as the symmetric spin splitting. 
The significant observations are that the above combinations of hopping parameters are composed so that the product of the hoppings belongs to the same irreducible representation as the spin-independent part of the AFM mean field.
In other words, the above combinations indicate the effective hoppings coupling to the AFM mean field; 
see the Supplemental Material~\cite{suppl} in detail for the analysis of the expansion method using MultiPie\cite{Kusunose_PhysRevB.107.195118}.
Such important model parameters are also extracted by using the expansion method for the magnetization~\cite{Hayami_PhysRevB.101.220403, Hayami_PhysRevB.102.144441, Oiwa_doi:10.7566/JPSJ.91.014701, Yatsushiro_PhysRevB.105.155157, Hayami_PhysRevB.106.024405, Kirikoshi_PhysRevB.107.155109}.

To summarize, we have investigated the linear and nonlinear PME under the $d$-wave and $g$-wave AMs. 
By examining the tight-binding model with the two-dimensional tetragonal collinear AFM orderings under the $p4mm$ and $p4gm$ space groups, we have found that the net magnetization is induced by the first and second order of strain, respectively; the induced magnetization for the former $p4mm$ model is characterized by $u_{xy}$, while that for the latter $p4gm$ model is characterized by $u_{xy}(u_{xx}-u_{yy})$.
Although the $g$-wave spin order observed in previously discovered materials exhibits a different symmetry compared to that in the present study, the coupling between the second order of strain and spin is a common feature of $g$-wave order, making the argument discussed in this Letter essentially the same and valid. 
In addition, present results indicate that the nonlinear PME in the $g$-wave AMs is attributed to the nonrelativistic magnetic-order driven effect, whereas the linear one in the $g$-wave AMs is owing to the effect of the SOC by considering the corresponding magnetic point group.
Thus, our results indicate that the PME is a good phenomenon for investigating the nature of not only $d$-wave AMs but also $g$-wave AMs.

\begin{acknowledgments} 
This research was supported by JSPS KAKENHI Grants Numbers JP21H01037, JP22H00101, JP22H01183, JP23H04869, JP23K03288, JP23K20827, and by JST CREST (JPMJCR23O4) and JST FOREST (JPMJFR2366). 
\end{acknowledgments}

\bibliography{letter_ver.9.bib}

\begin{thebibliography}{10}
\bibitem{observedPME}
The experimental observation of the linear PME in MnTe is supposed to be the effect of the SOC. The nonrelativistic contribution arises in the nonlinear PME, as discussed in the present study.
\end{thebibliography}
\bibliographystyle{jpsj}

\end{document}